\documentclass{PoS}

\newcommand{\bq}{\mbox{\boldmath $q$}}
\newcommand{\bk}{\mbox{\boldmath $k$}}
     
\title{Diffractive pQCD mechanism of exclusive production of $W^+ W^-$ pairs in proton-proton collisions}

\ShortTitle{Diffractive pQCD mechanism of exclusive production of $W^+ W^-$ pairs in $pp$ collisions}

\author{\speaker{Piotr Lebiedowicz}%
        \thanks{This work was supported in part by the MNiSW grant No. DEC-2011/01/N/ST2/04116.}\\
Institute of Nuclear Physics PAN, PL-31-342 Cracow, Poland\\
E-mail: \email{Piotr.Lebiedowicz@ifj.edu.pl}} 

\author{Roman Pasechnik\\
Department of Astronomy and Theoretical Physics, Lund University, SE-223 62 Lund, Sweden\\
E-mail: \email{Roman.Pasechnik@thep.lu.se}}

\author{Antoni Szczurek\\
University of Rzesz\'ow, PL-35-959 Rzesz\'ow, Poland, and\\
Institute of Nuclear Physics PAN, PL-31-342 Cracow, Poland\\
E-mail: \email{Antoni.Szczurek@ifj.edu.pl}}

\abstract{
We present a study of central exclusive production of $W^+W^-$ pairs 
in proton-proton collisions at the LHC.
We compare the contribution of the $\gamma \gamma \to W^+ W^-$ mechanism
with a new mechanism of exclusive diffractive production
through the $gg \to W^+ W^-$ subprocess
with intermediate virtual Higgs boson and quark box diagrams.
The amplitude for the latter process is expressed in terms
of the off-diagonal unintegrated gluon distribution functions.
Several observables related to this process are calculated.
The phase space integrated diffractive contribution when separated
is only a small fraction of fb compared to 115.4 fb 
of the $\gamma\gamma$-contribution without absorption.
This opens a possibility of efficient searches for anomalous boson
$\gamma W^+ W^-$ and $\gamma \gamma W^+ W^-$ couplings due to new physics
beyond Standard Model.}

\FullConference{Sixth International Conference on Quarks and Nuclear Physics\\
		April 16-20, 2012\\
		Ecole Polytechnique, Palaiseau, Paris}

\begin{document}

%---------------------------
\section{Introduction}
%---------------------------

Central exclusive production (CEP) processes $pp\to p + X + p$,
where $X$ stands for a centrally produced system separated from the
two very forward protons by large rapidity gaps,
provide a very promising way to investigate both QCD dynamics and 
new physics in hadron collisions.

We focus on central exclusive production of $W^+W^-$ pairs 
in high-energy proton-proton collisions \cite{LPS12}.
The $\gamma \gamma$-contribution for the purely exclusive production 
of $W^+ W^-$ was already considered in the literature.
The diffractive production and decay of Higgs boson into the $W^+W^-$ pair 
was discussed in Ref.~\cite{WWKhoze}, and the corresponding cross
section turned out to be smaller than that for 
the $\gamma\gamma$-contribution.
Provided this is the case, the $W^+W^-$ pair production signal would
be particularly sensitive to New Physics contributions in 
the $\gamma \gamma \to W^+ W^-$ subprocess \cite{royon,piotrzkowski}
and $pp \to pp W^+ W^-$ reaction is an 
ideal case to study experimentally $\gamma W^+ W^-$ and 
$\gamma \gamma W^+ W^-$ couplings
\footnote{Some more subtle aspects of the beyond Standard Model
anomalous couplings were discussed e.g. in \cite{MMN08}.}.

\begin{figure}[!h]
\begin{center}
\includegraphics[width=.35\textwidth]{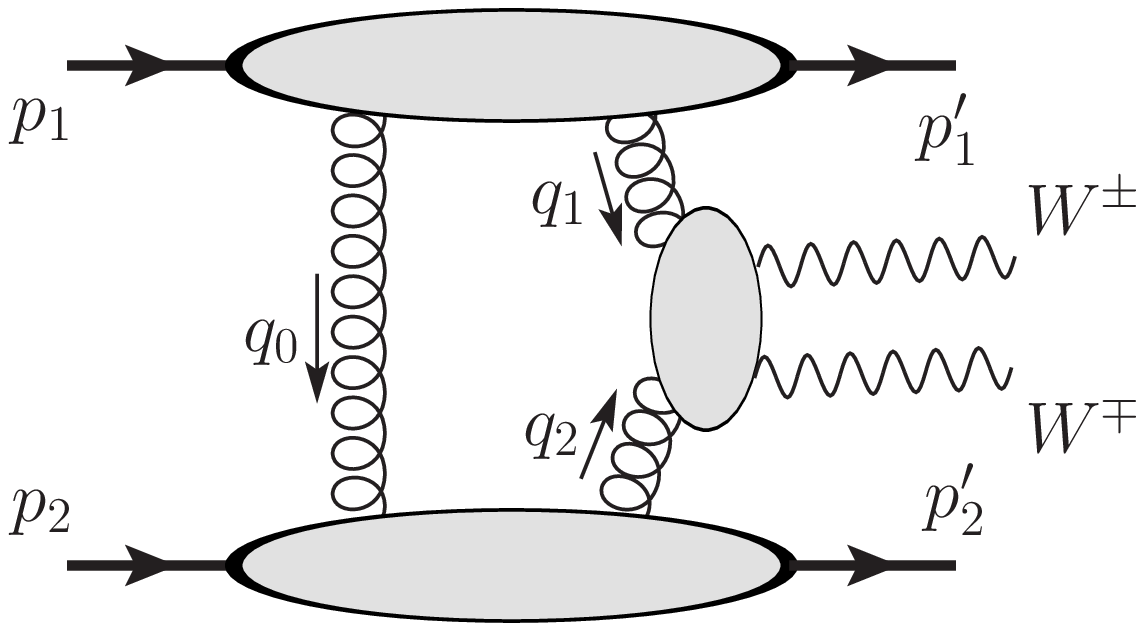} \qquad
\includegraphics[width=.5\textwidth]{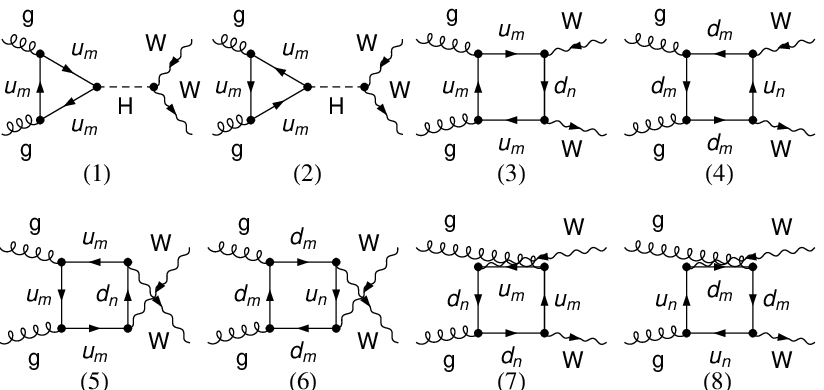}
\end{center}
\caption{
Diagram for the central exclusive $WW$ pair production in $pp$ collisions (left panel)
and representative diagrams of the hard subprocess $gg\to W^{\pm}W^{\mp}$ (right panel).}
\label{fig:WWCEP}
\end{figure}

The linear collider would be a good option to study the couplings of 
gauge bosons in the distant future.
For instance in Ref.\cite{NNPU} the anomalous coupling in locally 
SU(2) $\times$ U(1) invariant effective Lagrangian was studied. 
Other models also lead to anomalous gauge boson coupling. 
Similar analysis has been considered recently
for $\gamma \gamma \to Z Z$ \cite{Gupta:2011be}. 
These previous analyses strongly motivate our present detailed study 
on a competitive diffractive contribution.
The $pp\to pW^+W^-p$ process going through the diffractive QCD
mechanism with the $gg \to W^+W^-$ subprocess (shown in Fig.~\ref{fig:WWCEP}) 
naturally constitutes
a background for the exclusive electromagnetic 
$pp\to p(\gamma\gamma\to W^+W^-)p$ process.
We consider not only the mechanism with intermediate Higgs boson but
also box contributions never estimated in exclusive processes.
We discuss the interference effects within the Standard Model 
as potentially important irreducible
background for the $\gamma\gamma\to W^+W^-$ signal relevant for a
precision study of anomalous couplings. Thus, our numerical
estimates provide minimal limit for the central exclusive $WW$ production signal.

Corresponding measurements will be possible to perform at the ATLAS
detector with the use of very forward FP220 detectors \cite{royon}. 
In order to quantify to what extent the QCD mechanism competes with the
``signal'' from the $\gamma\gamma$ fusion, we calculate both
contributions and compare them as a function of several relevant phase space variables.

%----------------------------------------------------
\section{Formalism}
%----------------------------------------------------

%----------------------------------------------------
\subsection{Diffractive mechanism of exclusive $W^+W^-$ pair production}
%----------------------------------------------------
%In the central exclusive $W^+W^-$ production, triangle diagrams with
%$\gamma$ and $Z$ bosons in the intermediate state are suppressed due
%to the $J_z$ = 0 and parity selection rule for singlet gluon-gluon to 
%(virtual) photon transition strictly valid in the on-shell limit of 
%fusing gluons and Landau-Yang theorem for intermediate $Z$ boson.

We write the amplitude of the diffractive process, which at high energy
is dominated by its imaginary part, as
\begin{eqnarray} \label{ampl}
{\cal M}_{\lambda_+\lambda_-}(s,t_1,t_2) &\simeq&is\frac{\pi^2}{2}
\int d^2 {\bq}_{0\perp} V_{\lambda_+\lambda_-}(q_1,q_2,k_{+},k_{-})
\frac{f_g(q_0,q_1;t_1)f_g(q_0,q_2;t_2)}
{{\bq}_{0\perp}^2\,{\bq}_{1\perp}^2\,{\bq}_{2\perp}^2}\,,
\end{eqnarray}
where $\lambda_{\pm}=\pm 1,\,0$ are the polarisation states
of the produced $W^{\pm}$ bosons, respectively, $f_g(r_1,r_2;t)$ is
the off-diagonal unintegrated gluon distribution function (UGDF),
which depends on the longitudinal
and transverse components of both gluons momenta (see \cite{LPS12} for more details).
In the calculations we use the GJR NLO collinear gluon distribution \cite{GJR}
which allow to use very low $q_{\perp,cut}^{2} = 0.5$ GeV$^{2}$.

The momenta of intermediate gluons are given by Sudakov
decompositions in terms of the incoming proton four-momenta $p_{1,2}$
\begin{eqnarray}\nonumber
&&q_1=x_1p_1+q_{1\perp},\quad q_2=x_2p_2+q_{2\perp},\quad 0<x_{1,2}<1,\\
&&q_0=x'p_1-x'p_2+q_{0\perp}\simeq q_{0\perp},\quad x'\ll x_{1,2},
\label{moms}
\end{eqnarray}
where $x_{1,2}$, $x'$ are the longitudinal momentum fractions for
active (fusing) and color screening gluons, respectively, such that
$q_{\perp}^2\simeq -|\bq_{\perp}|^2$.
In the forward proton scattering limit, we have
\begin{eqnarray}
t_{1,2}=(p_{1,2}-p'_{1,2})^2\simeq{p'}^2_{1,2\perp}\to0,\quad
q_{\perp} \equiv q_{0\perp} \simeq -q_{1\perp} \simeq q_{2\perp} \,.
\label{forward}
\end{eqnarray}
%
%The QCD factorisation of the process at the hard scale $\mu_F$ is
%provided by the large invariant mass of the $WW$ pair $M_{WW}$, i.e.
%
%\begin{eqnarray}\label{sx1x2}
%\mu_F^2\equiv s\,x_1x_2\simeq M_{WW}^2\,.
%\end{eqnarray}
%
It is convenient to introduce the Sudakov expansion for $W^{\pm}$ boson momenta
\begin{eqnarray}
k_+=x_1^+ p_1+x_2^+ p_2+k_{+\perp},\quad
k_-=x_1^- p_1+x_2^- p_2+k_{-\perp}
\end{eqnarray}
leading to
\begin{eqnarray}\label{xqq}
x_{1,2}=x_{1,2}^+ + x_{1,2}^-,\quad
x_{1,2}^+=\frac{m_{+\perp}}{\sqrt{s}}e^{\pm y_+},\quad
x_{1,2}^-=\frac{m_{-\perp}}{\sqrt{s}}e^{\pm y_-},\quad
m_{\pm\perp}^2=m_W^2+|{\bk}_{\pm\perp}|^2\,,
\end{eqnarray}
in terms of $W^{\pm}$ rapidities $y_{\pm}$ and transverse masses
$m_{\pm\perp}$. For simplicity, in actual calculations we work in
the forward limit given by Eq.~(\ref{forward}), which implies that
${\bk}_{+\perp}=-{\bk}_{-\perp}$.

The gauge-invariant $gg\to W_{\lambda_+}^+ W_{\lambda_-}^-$
hard subprocess amplitude
$V_{\lambda_+\lambda_-}(q_1,q_2,k_{+},k_{-})$
is given by the light cone projection
\begin{eqnarray}\label{GIproj}
V_{\lambda_+\lambda_-}=
n^+_{\mu}n^-_{\nu}V_{\lambda_+\lambda_-}^{\mu\nu}=
\frac{4}{s}
\frac{q^{\nu}_{1\perp}}{x_1}
\frac{q^{\mu}_{2\perp}}{x_2}
V_{\lambda_+\lambda_-,\mu\nu},\quad
q_1^{\nu}V_{\lambda_+\lambda_-,\mu\nu}=
q_2^{\mu}V_{\lambda_+\lambda_-,\mu\nu}=0\,,
\end{eqnarray}
where $n_{\mu}^{\pm} = p_{1,2}^{\mu}/E_{p,cms}$ and the
center-of-mass proton energy $E_{p,cms} = \sqrt{s}/2$. We adopt the
definition of gluon transverse polarisation vectors proportional to
the transverse gluon momenta $q_{1,2 \perp}$, i.e. $\epsilon_{1,2} \sim q_{1,2 \perp} /
x_{1,2}$. The helicity matrix element in the previous expression reads
%$\epsilon(k_{+},\lambda_+)$ and $\epsilon(k_{-},\lambda_-)$
%with helicities $\lambda_+, \lambda_-$ and momenta $k_+, k_-$, respectively,
%
\begin{eqnarray}
V_{\lambda_+\lambda_-}^{\mu\nu}(q_1,q_2,k_{+},k_{-})=
\epsilon^{*,\rho}(k_+,\lambda_+)
\epsilon^{*,\sigma}(k_-,\lambda_-)V_{\rho\sigma}^{\mu\nu}\,,
\label{Vepsilon}
\end{eqnarray}
in terms of the Lorentz and gauge invariant $2\to2$ amplitude
$V_{\rho\sigma}^{\mu\nu}$ and $W$ boson polarisation vectors
$\epsilon(k,\lambda)$, which can be defined easily in the
$pp$ center-of-mass frame with $z$-axis along the proton beam (see \cite{LPS12}).
The matrix element $V_{\lambda_{+},\lambda_{-}}$ contains twice the
strong coupling constant $g_s^2 = 4 \pi \alpha_{s}$.
We take the running coupling constant
$\alpha_s(\mu_{hard}^2=M_{WW}^2)$ which depends on the invariant
mass of $WW$ pair taken as a hard renormalisation scale of the process.
In the calculations we use the reduced form of the
four-body phase space \cite{LPS12, LS2010}.

%--------------------------------------------------------------------------
\subsection{Electromagnetic $\gamma \gamma \to W^+ W^-$ mechanism}
%--------------------------------------------------------------------------

In the Weizs\"acker-Williams approximation,
the total cross section for the $pp \to pp (\gamma \gamma \to W^+ W^-)$
can be written as in the parton model
\begin{equation}
\sigma = \int d x_1 d x_2 \, f_1^{WW}(x_1) \, f_2^{WW}(x_2) \,
\hat{\sigma}_{\gamma \gamma \to W^+ W^-}(\hat s) \,,
\label{EPA}
\end{equation}
where we take the Weizs\"acker-Williams equivalent photon fluxes of protons from Ref.~\cite{DZ}.
The elementary tree-level cross
section for the $\gamma \gamma \to W^+ W^-$ subprocess can be written in the
compact form in terms of the Mandelstam variables (see e.g. Ref.~\cite{DDS95}).
To calculate differential distributions the following parton formula
can be used
\begin{equation}
\frac{d\sigma}{d y_+ d y_- d^2 p_{W\perp}} = \frac{1}{16 \pi^2 {\hat s}^2}
\, x_1 f_1^{WW}(x_1) \, x_2 f_2^{WW}(x_2) \,
\overline{ | {\cal M}_{\gamma \gamma \to W^+ W^-}(\hat s, \hat t, \hat u)
  |^2} \, ,
\label{EPA_differential}
\end{equation}
where the momentum fractions of the fusing gluons $x_{1,2}$ are defined in Eq.~(\ref{xqq}).

%--------------------------------------------------------------------------
\subsection{Inclusive production of $W^+W^-$ pairs (gluon-gluon fusion)}
%--------------------------------------------------------------------------
%For a test and for a comparison we also consider a gluon-gluon contribution
%to the inclusive cross section.
In the lowest order of pQCD the inclusive cross section can be written as \cite{LPS12}
\begin{equation}
\frac{d \sigma}{d y_+ d y_- d^2 p_{W\perp}} = \frac{1}{16 \pi^2 {\hat s}^2}
x_1 g(x_1, \mu_F^2) x_2 g(x_2,\mu_F^2)
\overline{| {\cal M}_{gg \to W^+ W^-}(\lambda_1, \lambda_2, \lambda_+,
  \lambda_-) |^2} \, .
\label{inclusive_cs}
\end{equation}
%
%The corresponding matrix elements have been discussed in the
%literature in detail \cite{inclusive}. The distributions in rapidity
%of $W^+$ ($y_+$), rapidity of $W^-$ ($y_-$) and transverse momentum
%of one of them $p_{W\perp}$ can be calculated in a straightforward way from
%Eq.~(\ref{inclusive_cs}). The distribution in invariant mass can be
%then obtained by an appropriate binning.

%--------------------------------------------------------------------------
\section{Results}
%--------------------------------------------------------------------------
%-----------------------------------------------------------------------------
\begin{figure}[!h]
\begin{center}
\includegraphics[width=4.9cm]{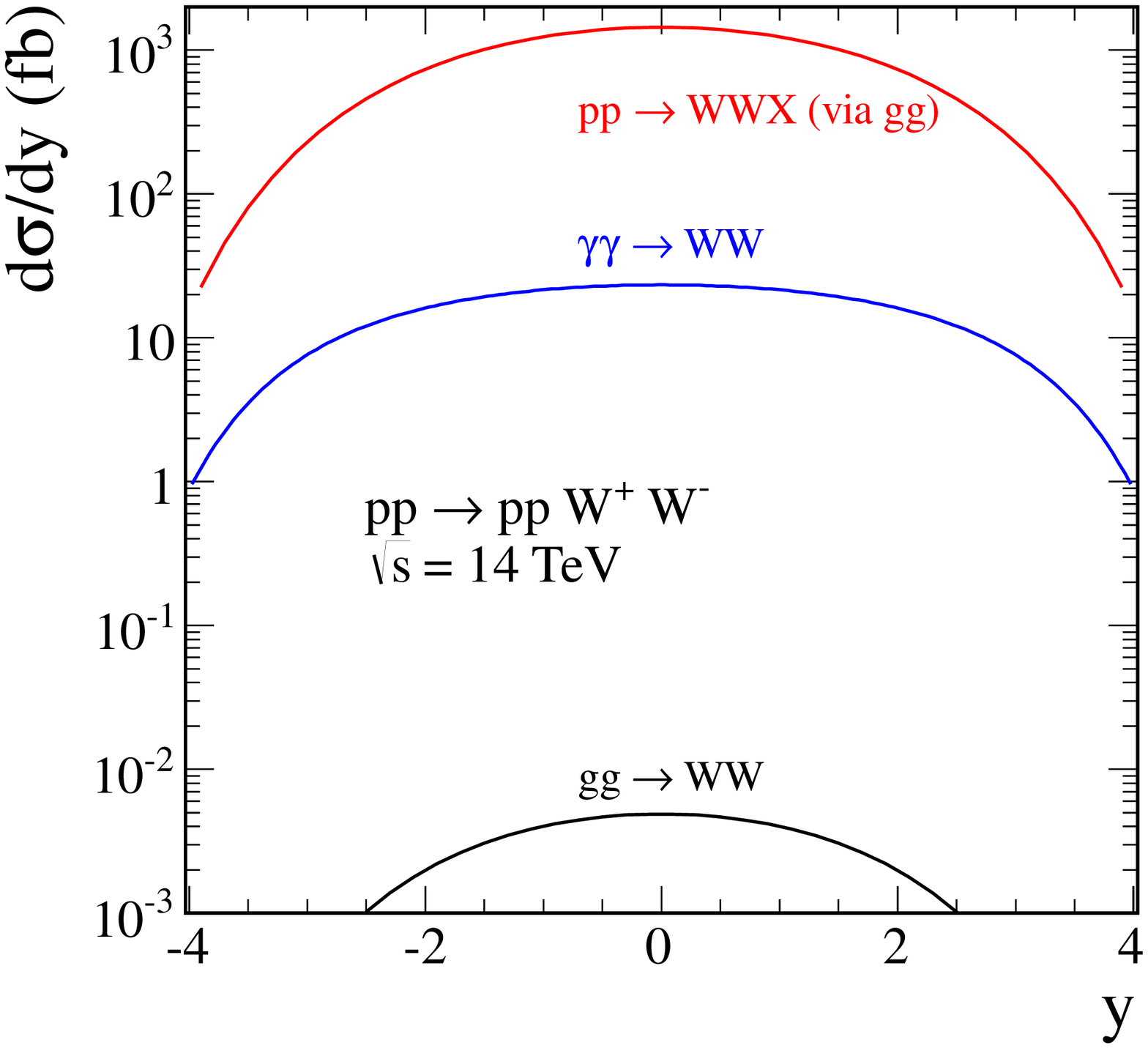}
\includegraphics[width=4.9cm]{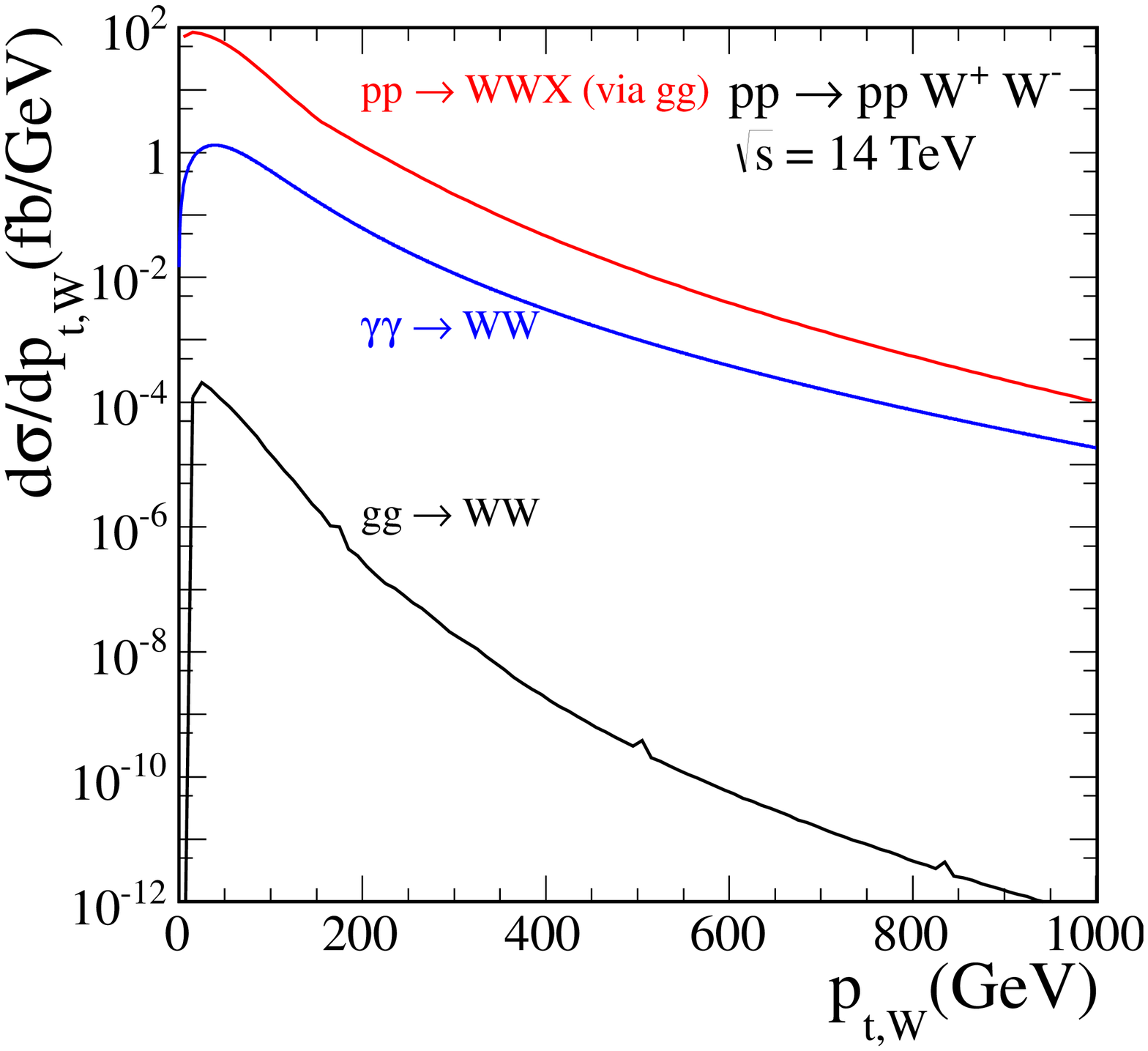}
\includegraphics[width=4.9cm]{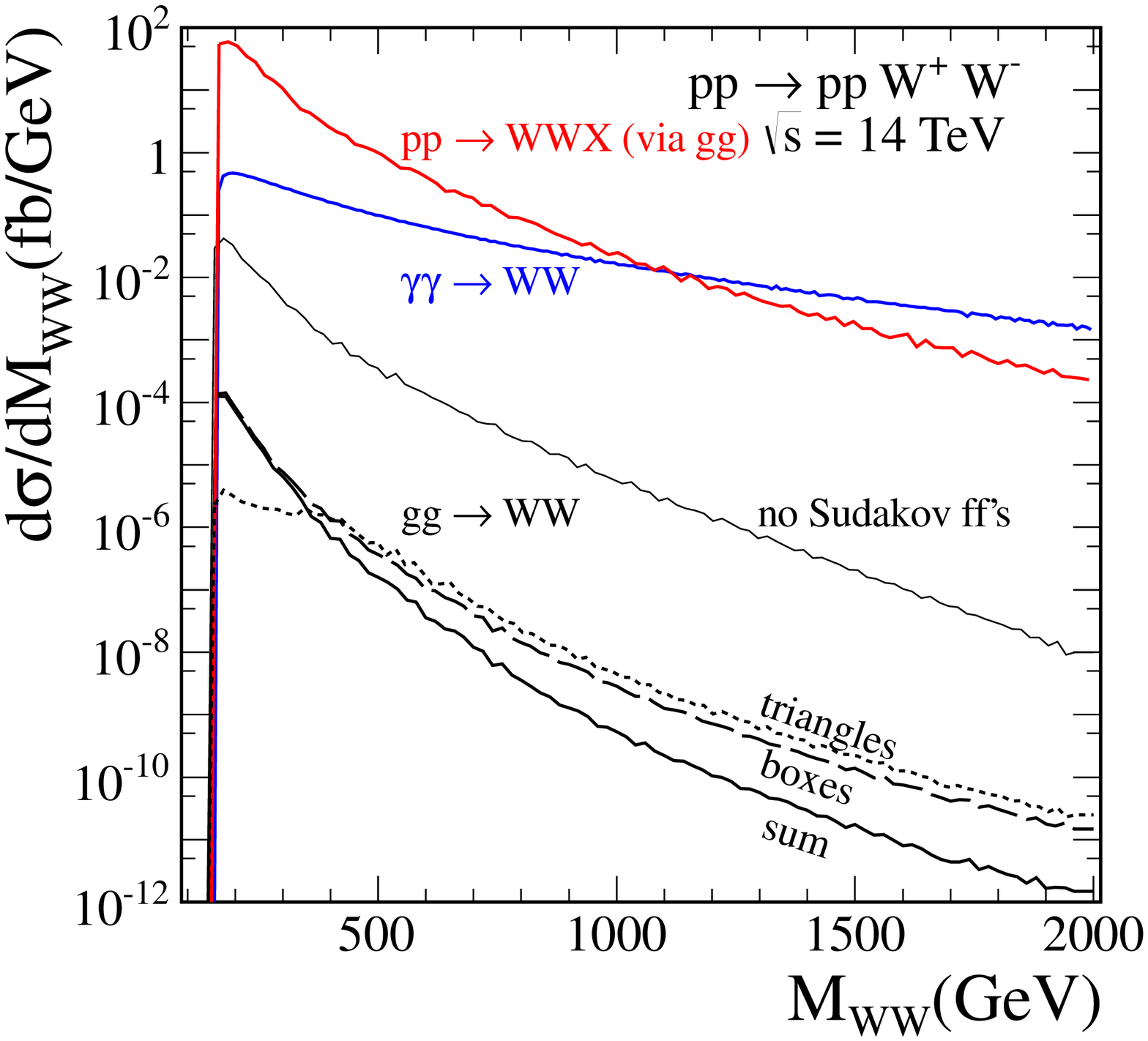}
\end{center}
   \caption{
\small Distribution in $W$ bosons rapidity (left panel), transverse momentum of one of the $W$ bosons (central panel) and $W^+W^-$ invariant mass (right panel).
The diffractive contribution is shown by the bottom line
while the $\gamma \gamma \to W^+ W^-$ contribution by the middle line.
The top line corresponds to the inclusive
gluon-initiated $pp \to W^+W^-X$ component.
On the right panel the result when the Sudakov form factor is
put to one is shown.
}
\label{fig:results}
\end{figure}
%------------------------------------------------------------------------------
The main results emerging from our analysis are presented in Fig.~\ref{fig:results}.
We compare distributions of $W$ boson for 
the electromagnetic $\gamma \gamma \to W^+ W^-$ 
and diffractive $gg \to W^+ W^-$ mechanisms. For a reference, we show also
inclusive cross section ($g g \to W^+ W^-$ contribution only) which
is five orders of magnitude bigger than the exclusive counterpart.
The two-photon induced contribution is almost three orders of magnitude larger than the
diffractive contribution, in which all polarization components for
$W^+$ and $W^-$ have been included.
It is completely opposite than for 
$p p \to p H p$, $p p \to p Q\bar Q p $ \cite{MPS_bbbar} or $p p \to p M p$ 
(e.g. light/heavy quarkonia production \cite{chic,chic_LPS,LKRS10}) CEP processes.
The standard relative suppression is due to soft gap survival probability factor
($S_g \sim$ 0.03 for diffractive contribution versus $S_g \sim$ 1 
for two-photon contribution), and due to a suppression by
the Sudakov form factor (see e.g. \cite{LPS12}) calculated at very large scales, here at
$\mu_{hard} = M_{WW}$. The main difference compared to other cases is that
in the diffractive case the leading contribution comes from loop
diagrams while in the two-photon case already from tree level diagrams.

The distribution in $W^+$ ($W^-$) transverse momentum for exclusive diffractive production is much steeper than that for the electromagnetic contribution.
The diffractive contribution peaks at $p_{t,W} \sim$ 25 GeV
while for the $\gamma \gamma \to W^+ W^-$ mechanism
the maximum is at $p_{t,W} \sim$ 40 GeV.
The exclusive cross section for $\gamma \gamma$-contribution
is at large $p_{t,W} \sim$ 1 TeV smaller
only by one order of magnitude than the inclusive $gg \to W^{+}W^{-}$ component.
The situation could be more favorable if New Physics would be at the game \cite{royon}.

The distribution for the diffractive component
drops quickly with the $M_{WW}$ invariant mass. As can be seen from
the figure, the Sudakov form factor lowers the cross section by a
large factor. The larger $M_{WW}$ the larger the damping. 
We show the full result (boxes + triangles) and the result
with boxes only which would be complete if the Higgs boson
does not exist. At high invariant masses, the interference of boxes
and triangles decreases the cross section. The distribution for the
$\gamma \gamma$-component drops very slowly with $M_{WW}$ and
at $M_{WW} >$ 1 TeV the corresponding cross section is even bigger than
the $gg \to W^+ W^-$ component to inclusive production of $W^+ W^-$ pairs.

%--------------------------
\section{Conclusions}
%--------------------------
We have perform a complete calculation of the QCD diffractive contribution to the exclusive
$p p \to p W^+ W^- p$ process for the first time in the literature
with the full one-loop (leading order) $gg\to W^+W^-$ matrix element. Two mechanisms
have been considered. First mechanism is a virtual (highly off-shell)
Higgs boson production and its subsequent transformation into real $W^+ W^-$ pair. 
Second mechanism relies on the formation of
intermediate quark boxes.
We have calculated corresponding amplitudes using computer
program package {\tt FormCalc} \cite{FC}. We have made a first estimate of
the cross section using amplitudes in the forward limit ``corrected''
off-forward using a simple exponential (slope dependent) extrapolation.
We have shown that extra box diagrams,
even though they are larger than the resonant
diagrams, constitute a negligibly small background for a precision
study of anomalous couplings.

Differential distributions in the $W^{\pm}$ transverse momentum,
rapidity and $W^+ W^-$ pair invariant mass have been calculated and
compared with corresponding distributions for the discussed in the
literature $\gamma \gamma \to W^+ W^-$ mechanism.
The contribution of triangles with the intermediate Higgs boson
turned out to be smaller
than the contribution of boxes taking into account recent
very stringent limitations on Higgs boson mass from the Tevatron and LHC data.
We have found that, in contrast to exclusive production of
Higgs boson or dijets, the two-photon fusion dominates over the
diffractive mechanism for small four-momentum transfers squared in
the proton lines as well as in a broad range of $W^+W^-$ invariant masses, in particular, for large $M_{WW}$.
Estimated theoretical uncertainties cannot disfavor this statement.
The large $M_{WW}$ region is damped in the diffractive model via
scale dependence of the Sudakov form factor.

One could focus on the diffractive contribution by imposing lower
cuts on $t_1$ and/or $t_2$ using very forward detectors
on both sides of the interaction point at distances of
220 m and 420 m as planned for future studies at ATLAS and CMS. 
The corresponding cross section is, however, expected to be extremely low.

The unique situation of the dominance of the $\gamma \gamma \to W^+
W^-$ contribution over the diffractive one opens a possibility of
independent tests of the Standard Model as far as the triple-boson
$\gamma W W$ and quartic-boson $\gamma \gamma W W$ coupling
is considered. It allows also stringent tests of some Higgsless
models as discussed already in the literature (see e.g. Ref.~\cite{royon}).

\end{document}